\documentclass[
 reprint,
 amsmath,amssymb,superscriptaddress,
 aps,
 prl
]{revtex4-1}

\usepackage{graphicx}
\usepackage{dcolumn}
\usepackage{bm}
\usepackage{color}
\usepackage{ulem}
\usepackage{float}
\usepackage{textcomp}
\restylefloat{table}
\usepackage{verbatim}

\begin{document}

\title{Splitting of the Fermi Contour of Quasi-2D Electrons in Parallel Magnetic Fields}
\date{\today}

\author{M. A.\ Mueed}
\author{D.\ Kamburov}
\author{M.\ Shayegan}
\author{L. N.\ Pfeiffer}
\author{K. W.\ West}
\author{K. W.\ Baldwin}
\affiliation{ Department of Electrical Engineering, Princeton University, Princeton, New Jersey 08544, USA}
\author{R.\ Winkler}
\affiliation{Department of Physics, Northern Illinois University, DeKalb, Illinois 60115, USA}
\affiliation{Materials Science Division, Argonne National Laboratory, Argonne, Illinois 60439, USA}

\begin{abstract}
In a quasi two-dimensional electron system with non-zero layer thickness, a parallel magnetic field can couple to the out-of-plane electron motion and lead to a severe distortion and eventual splitting of the Fermi contour.  Here we directly and quantitatively probe this evolution through commensurability and Shubnikov-de Haas measurements on electrons confined to a 40-nm-wide GaAs (001) quantum well. We are able to observe the Fermi contour splitting phenomenon, in good agreement with the results of semi-classical calculations. Experimentally we also observe intriguing features, suggesting magnetic-breakdown-type behavior when the Fermi contour splits.
\end{abstract}

\maketitle

\begin{figure*}
\includegraphics[width=1\textwidth]{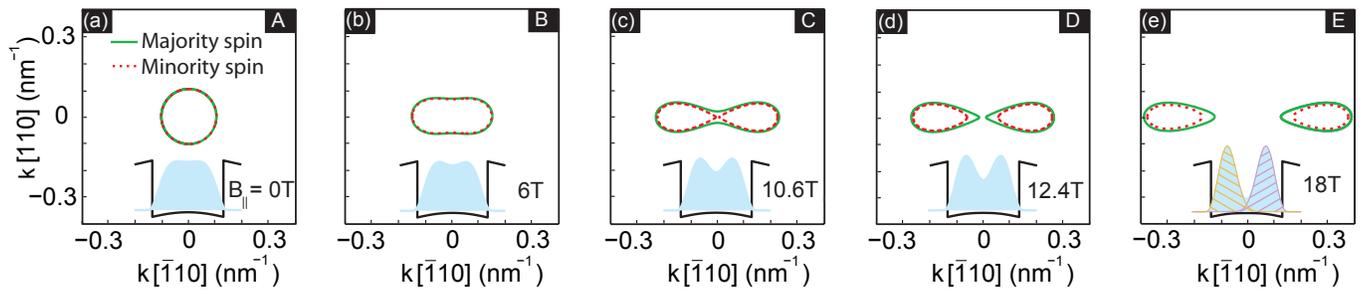}
\caption{\label{fig:Fig1} (color online) (A)-(E) Calculated Fermi contours for a 2DES with density $n=1.75\times10^{11}$ cm$^{-2}$ confined to a 40-nm-wide GaAs QW. $B_{||}$ is applied along [110]. Insets show the corresponding charge distributions in light blue. Inset of (E) also shows the charge distribution from $k[\overline{1}10]>0$ (shaded pink region) and $k[\overline{1}10]<0$ (shaded yellow region) states. }
\end{figure*}

\begin{figure}
\includegraphics[width=.48\textwidth]{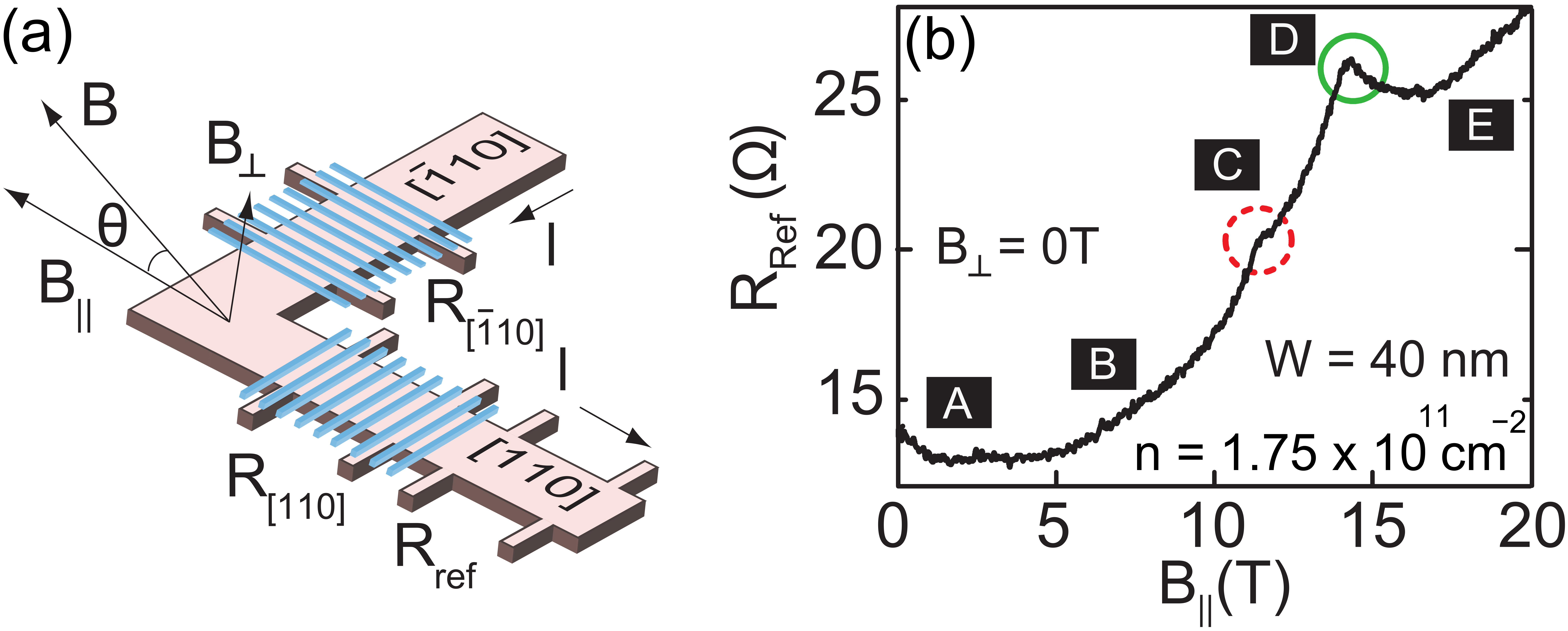}
\caption{\label{fig:Fig1} (color online) (a) Sample schematics. The electron-beam resist grating covering the surface of each Hall bar arm is shown as blue stripes. Part of the [110] arm is left unpatterned as a reference region. (b) $B_{||}$-induced magnetoresistance for $R_{ref}$. (A)-(E) mark the $B_{||}$ values that correspond approximately to the calculations of Fig. 1.}
\end{figure}

\begin{figure}
\includegraphics[width=.461\textwidth]{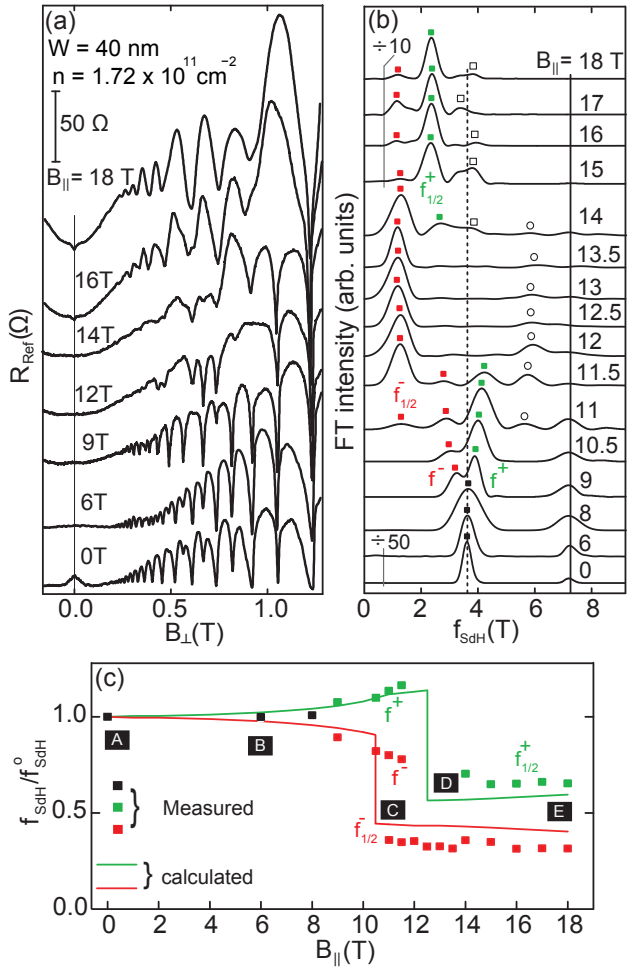}
\caption{\label{fig:Fig1} (color online)  (a) SdH oscillations measured in the reference region of the Hall bar as $B_{||}$ increases; traces are shifted vertically for clarity. (b) FT spectra of the SdH oscillations. The dotted and solid black lines show the expected positions of the spin-unresolved and spin-resolved FT peaks at $B_{||}=0$, respectively. The FT signal to the left of the vertical lines indicated by $\div10$ and $\div50$ is affected by the Hamming window used in the Fourier analysis and is shown suppressed. (c) Summary of the FT peak positions normalized to $f^{o}_{SdH}$, the frequency at $B_{||}=0$. Closed squares represent the measured frequencies. The frequencies predicted by the calculations for the spin subbands are shown as green and red lines. (A)-(E) mark the $B_{||}$ values from Fig. 1 calculations.}
\end{figure}

\begin{figure}
\includegraphics[width=.48\textwidth]{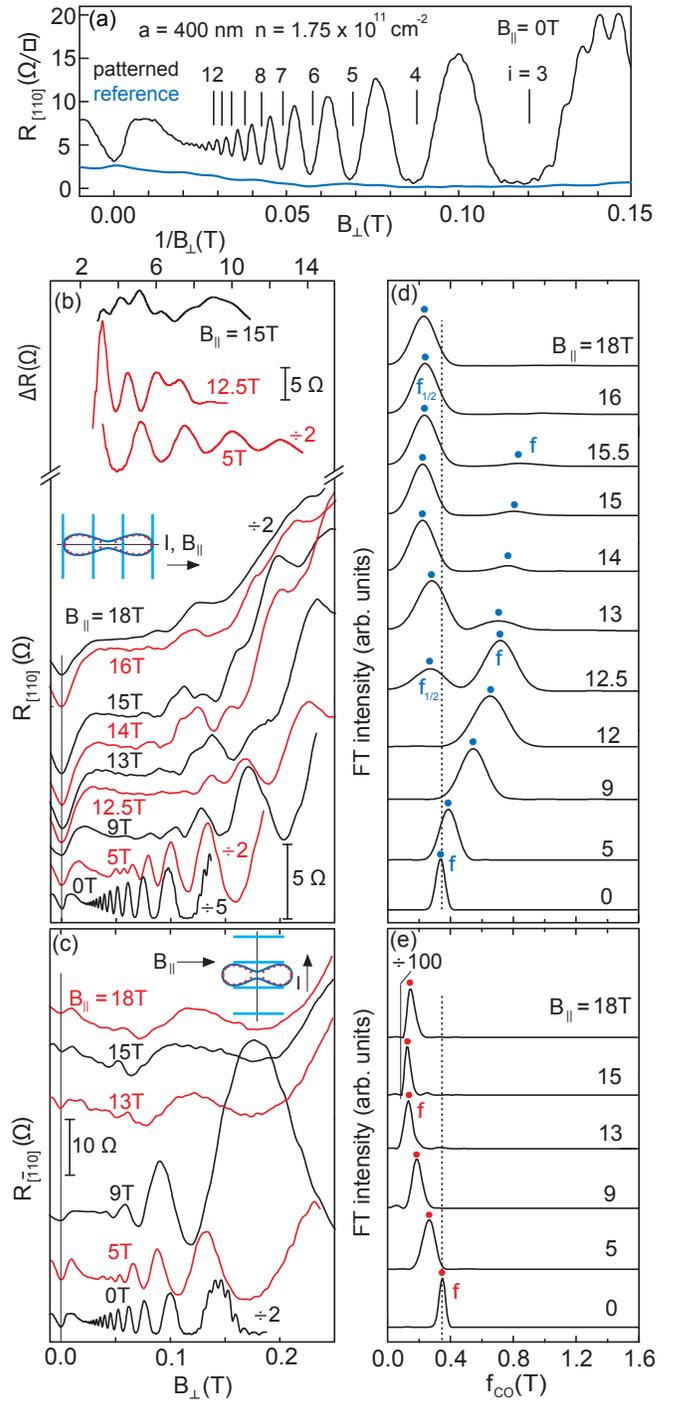}
\caption{\label{fig:Fig1} (color online) (a) COs along [110] from the patterned region. Vertical lines mark the expected positions of the COs resistance minima. 
The corresponding trace from the unpatterned reference region clearly shows no COs. (b), (c) COs from the patterned regions of the L-shaped Hall bar along [110] and $[\overline{1}10]$ for different values of $B_{||}$. In order to do Fourier analysis, we extract the oscillatory part of these traces as a function of $1/B_{\perp}$ by subtracting the background resistance. Panel (b) also includes three examples of this process. Traces are shifted vertically for clarity. (d), (e) Normalized FT spectra of the COs data shown in (a) and (b), respectively. The vertical dotted lines mark $f^{0}_{CO}$ (see text).}
\end{figure}

\begin{figure}
\includegraphics[width=.46\textwidth]{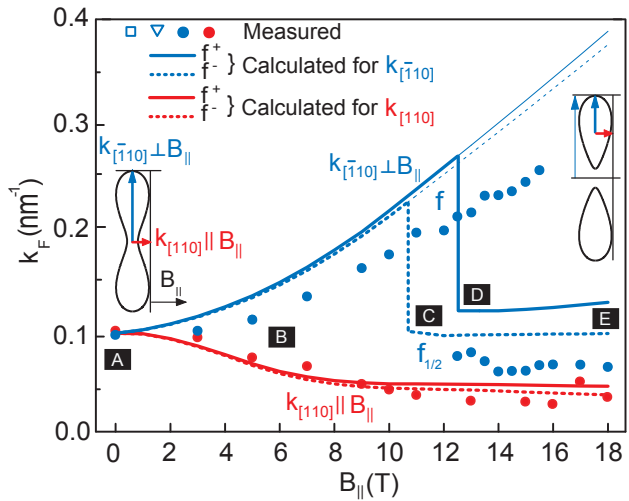}
\caption{\label{fig:Fig1} (color online) Summary of the Fermi wave vectors ($k_{F}$) deduced from the positions of the COs' FT spectra for the two Hall bar arms. Blue and red symbols represent the experimental data for $k_{F}$$\perp$$B_{||}$ and $k_{F}||B_{||}$, respectively. (A)-(E) mark the $B_{||}$ values from the calculations of Fig. 1.}
\end{figure}

In a strictly two-dimensional electron system (2DES) with zero layer thickness, the electron in-plane motion is unaffected by a parallel magnetic field ($B_{||}$). However, for a quasi-2DES, such as electrons in a quantum well (QW) with finite width, $B_{||}$ can couple to electrons' out-of-plane motion, thus also affecting their in-plane motion. This can have profound consequences. For example, the Fermi contour, which is circular in an isotropic system such as the 2DES in GaAs QWs, becomes severely distorted by $B_{||}$ and could even split into two, tear-drop shaped contours if $B_{||}$ is sufficiently strong (Fig. 1). $B_{||}$ also causes spin-polarization, leading to the formation of two distinct Fermi contours with different enclosed areas for different spins. Therefore, the spin-degenerate Fermi contour at $B=0$ could split into \textit{two pairs} of smaller contours in the presence of a large $B_{||}$ (Fig. 1). Since the shape of Fermi contour, when rotated by $90^{\circ}$, reflects that of the cyclotron orbit in real space \cite{Ashcroft.SSP.1976}, its evolution as a function of $B_{||}$ corresponds to the evolution of the electron trajectory. Understanding this $B_{||}$-induced Fermi contour splitting is of fundamental importance, specially for spintronic devices where application of $B_{||}$ is often used for spin-polarization \cite {footnote0,Rokhinson.PRL.2004}. 

Several transport studies on 2DESs, confined to coupled double- \cite{Eisenstein.PRB.1991,Kurobe.PRB.1994,Simmons.PRL.1994,Simmons.PRB.1995,Harff.PRB.1997,Blount.PRB.1998} and triple-QW systems \cite{Lay.PRB.1995} have previously explored the Fermi contour splitting. In these studies, features such as kinks in the $B_{||}$-induced magnetoresistance and inter-layer tunneling were associated with the splitting. Electrons in very wide single QWs, which are essentially bilayer systems, also produced similar results \cite{Jungwirth.PRB.1997}. Compared to the earlier works, our study here incorporates the following novelties: (1) The 2DES is confined to a single QW with a \textit{single-layer-like} charge distribution at $B_{||}=0$ (see inset of Fig. 1(A)). (2) We probe the splitting of the Fermi contour for both \textit{spin} species via Shubnikov-de Haas (SdH) oscillations. (3) We use measurements of commensurability oscillations (COs), also known as Weiss oscillations \cite{Weiss.EurophysL.1989}, to directly map out the Fermi contour and capture its distortion and the eventual splitting. 

Our sample, grown via molecular beam epitaxy, is a 40-nm-wide, GaAs (001) QW which is located 190 nm under the surface. The QW is flanked on each side by 95-nm-thick Al$_{0.24}$Ga$_{0.76}$As spacer layers and Si $\delta$-doped layers. The 2DES density is $n=1.75\times10^{11}$ cm$^{-2}$, and the mobility is $\sim 20\times$$10^{6}$ cm$^{2}$/Vs. We fabricated a strain-inducing superlattice with a period $a=400$ nm on the surface of our sample, an L-shaped Hall bar (Fig. 2 (a)). The superlattice, made of negative electron-beam resist, modulates the potential through the piezoelectric effect in GaAs \cite{Kamburov.PRB.2012,Kamburov2.PRB.2012,Skuras.APL.1997,Endo.PRB.2000,Kamburov.PRB.2013}. For $B_{||}$-dependent measurements, we first apply a large $B$-field in the plane along [110]. The sample, mounted on a single-axis tilting stage, is then slowly rotated around $[\overline{1}10]$ using a computer-controlled, brushless DC motor to introduce a small component of the field perpendicular to the 2D plane. This $B_{\perp}$ induces SdH oscillations in the unpatterned reference and COs in the modulated regions of the Hall bar. (Note that $B_{||}\cong{B}$ because $B_{\perp}\ll{B}$.) We pass current along the L-shaped Hall bar and measure the longitudinal resistances simultaneously for both arms. $B_{\perp}$ is extracted from a linear fit of the Hall resistance measured in the reference region. All measurements are carried out at 300 mK.

Figures 1(A)-(E) highlight the key points of our study. The Fermi contours are derived from calculations based on an $8\times8$ Kane Hamiltonian with no adjustable parameters \cite{Winkler.Springer.2003}. We include $\textbf{\textit{B}}_{||}=(B_{x}, B_{y}, 0)$ via the vector potential $\textbf{\textit{A}}(z)=(zB_{y}, -zB_{x}, 0)$ so that the in-plane canonical momentum
$\textbf{\textit{k}}=(k_{x}, k_{y}, 0)$ remains a good quantum number. The occupied states enclosed by the Fermi contours define the charge density from which the Hartree potential is derived self-consistently. As seen in Fig. 1(A), the Fermi contours of the two spin-subbands, which are identical for $B_{||}$=0, start out circular \cite{footnote1}. As $B_{||}$ increases, the 2DES becomes progressively more spin-polarized, thus two distinct spin Fermi contours are formed. The application of $B_{||}$ along [110], also elongates and splits the contours in the $[\overline{1}10]$ direction. Figures 1(C) and 1(D) illustrate the splitting of the minority and majority spin contours, respectively. The split contours separate even further along [110] under stronger $B_{||}$ (Fig. 1(E)). The insets of Figs. 1(A)-(E) show how the charge distribution along the confinement direction gradually evolves from a single-layer into a bilayer, with each of the ``layers" corresponding to one part of the split Fermi contour (see Fig. 1(E) inset).

As an indication of Fermi contour splitting, we first present the $B_{||}$-magnetoresistance trace from the reference region of the Hall bar (Fig. 2(b)). The trace demonstrates non-monotonic transport behavior. In particular, there are two pronounced kinks at $\cong11$ T and $\cong14$ T, marked by red and green circles, respectively. The positions of these kinks agree well with the expected splitting of the minority and majority spin contours (see Figs. 1(C) and (D)). A kink in the $B_{||}$-magnetoresistance has been previously associated with the splitting of Fermi contour \cite {Blount.PRB.1998}. Here, we observe two kinks, suggesting a spin-dependent splitting of Fermi contours. 

We further investigate the splitting via SdH oscillations which directly probe the area enclosed by the Fermi contour. We expect that the splitting would be reflected as a jump in the SdH frequency. Figure 3(a) shows the SdH oscillations at different $B_{||}$ while the corresponding Fourier transforms (FTs) are shown in Fig. 3(b). For $B_{||}=0$ T, we observe two peaks, the stronger of which ($f_{SdH}^{0}=3.56$ T) is for the spin-unresolved SdH oscillations (marked by a dotted vertical gray line). The weaker peak at 7.08 T is very close to the value of $2f_{SdH}^{0}$ (marked by a solid vertical gray line) and corresponds to the spin-resolved oscillations \cite{footnote1_1}. Around $B_{||}=9$ T, the spin-unresolved peak splits, with the lower frequency peak $f^{-}$ (red square) corresponding to the electron density of the minority-spin-subband and the higher frequency peak $f^{+}$ (green square) to the majority-spin-subband. Then, starting at $B_{||}\cong11$ T, another low-frequency peak $f^{-}_{1/2}$ (red square) appears at approximately $f^{-}/2$, signaling the splitting of the minority-spin contour. The $f^{-}_{1/2}$ peak remains dominant between $B_{||}=$ 11.5 and 13.5 T where both $f^{+}$ and $f^{-}$ become very weak and essentially vanish. However, at $B_{||}\cong15$ T, another peak $f^{+}_{1/2}$ (green square) appears to the right of $f^{-}_{1/2}$ and becomes the dominant feature in the FT spectrum up to $B_{||}=18$ T. The sum of $f^{-}_{1/2}$ and $f^{+}_{1/2}$ is close to $f_{SdH}^{0}$ implying that $f^{+}_{1/2}$ originates from the split majority spin Fermi contour. We do not fully understand the origin of the weak peaks marked by open symbols in Fig. 3(b). They might stem from magnetic breakdown between the split contours \cite{Ashcroft.SSP.1976.MB,Hu.PRB.1992}; similar phenomenon has been invoked to explain anomalous SdH frequencies seen in bilayer electron systems confined to double-QW samples in $B_{||}$ \cite{Harff.PRB.1997}.

We summarize, in Fig. 3(c), the results of the Fermi contour calculations and the measured SdH frequencies (red and green squares), normalized to $f_{SdH}^{0}$. The calculated frequencies for majority- and minority-spin contours, which are equal to the calculated Fermi contour areas multiplied by $h/(2\pi^{2}e)$, halve at $B_{||}$-values that mark the splitting of the respective contours. This jump in frequency reflects the fact that, for small $B_{||}$, the plotted curves are based on the areas enclosed by the unbroken Fermi contours, whereas for larger $B_{||}$, they are based on the areas of each of the split contours. There is good overall agreement between the measured and calculated SdH frequencies \cite{footnote2}. This suggests that SdH oscillations indeed show the spin-dependent splitting of Fermi contours, corroborating our interpretation of the two kinks observed in the $B_{||}$-magnetoresistance (Fig. 2(a)).

Having established the Fermi contour splitting through $B_{||}$-magnetoresistance and SdH oscillations, we now turn to COs data measured in the modulated regions of the Hall bar. The magnetoresistance trace of Fig. 4(a), taken as a function of purely $B_{\perp}$, is representative of such COs exhibiting pronounced minima at the electrostatic commensurability condition $2R_{C}/a=i-1/4$ \cite{Weiss.EurophysL.1989, Winkler.PRL.1989, Gerhardts.PRL.1989, Beenakker.PRL.1989, Beton.PRB.1990, Peeters.PRB.1992, Mirlin.PRB.1998}, where i = 1, 2, 3..., $R_{C}=k_{F}/eB_{\perp}$ is the real-space cyclotron diameter, and $a$ is the period of the potential modulation ($k_{F}$ is the Fermi wave-vector perpendicular to the current direction). The frequency of COs, $f_{CO}=2{\hbar}k_{F}/ea$, directly measures $k_{F}$. Note that the very high mobility of our sample leads to a large number of oscillations, up to $i\gtrsim12$. 

The magnetoresistance data from the [110] and $[\overline{1}10]$ Hall bar arms are shown in Figs. 4(b) and (c). In each figure, the bottom traces, taken in the absence of $B_{||}$, exhibit high quality COs. As $B_{||}$ is increased, there is an obvious change in the periodicity of COs which is better seen in the FT spectra of Figs. 4(d) and (e). The bottom FT spectrum from each of these figures exhibits a single peak whose position ($\simeq0.35$ T) is consistent with the commensurability frequency $f^{0}_{CO}=2{\hbar}k_{F}/ea=0.35$ T (dotted line) expected for a circular, spin-degenerate, Fermi contour with $k_{F}=\sqrt{2{\pi}n}$. With increasing $B_{||}$, this peak $f$ moves to higher frequencies in the FTs for the [110] Hall bar arm (Fig. 4(d)) and to lower frequencies in the $[\overline{1}10]$ arm (Fig. 4(e)), suggesting that the Fermi contour is getting elongated. However, at $B_{||}\cong12.5$ T, a new peak $f_{1/2}$ emerges at approximately $f/2$ in Fig. 4(d). This indicates that the elongated contour has split into two smaller ones. As $B_{||}$ is increased farther, $f_{1/2}$ develops to be the strongest feature of the FT spectra in Fig. 4(d). In contrast, $f$ becomes progressively less pronounced and vanishes at $B_{||}\cong16$ T. 

Figure 5 summarizes the values of $k_{F}$ extracted from the FT frequencies and also from the calculated Fermi contours for which we take the extrema along [110] and $[\overline{1}10]$ (see the left inset) and plot them with bold red and blue lines, respectively. However, after the splitting, $k_{F}$ along  $[\overline{1}10]$, defined as shown in Fig. 5 right inset (thin blue line with an arrow), is represented by a thin blue line. We also plot half the length of the major-axis along $[\overline{1}10]$ of the split contour (see the right inset) by a bold blue line. Qualitatively, the measured values of $k_{F}$ show good agreement with the calculations, suggesting that the peak $f_{1/2}$ comes from the split Fermi contour. Calculations (see Fig. 1) also show that the extreme sizes of the contours for the two spin species always remain very similar. This explains why, unlike the SdH oscillations data, COs do not resolve the two spin Fermi contours \cite{Kamburov.PRB.2013}. Another key point of Fig. 5 is that the elongation of the Fermi contour deduced from the COs data is smaller than what the calculations predict. A similar discrepancy was previously observed in other 2D electron and hole systems \cite{Kamburov.PRB.2013, Kamburov.PRB.2012, Smrcka.arxiv.2015}. 

Another noteworthy feature of the COs data is that, even after the Fermi contour splits at $B_{||}\cong12$ T, we appear to still follow $f$ up to $B_{||}\cong16$ T (marked by solid blue circles in Figs. 4(d) and 5). To explain this, we propose a magnetic breakdown like scenario \cite{Ashcroft.SSP.1976.MB,Hu.PRB.1992} where, even though the Fermi contour is split into two pieces, we still observe COs because a small portion of electrons jump between the split contours and complete the elongated orbit. In this context, magnetic breakdown in $k$-space implies that, in real space, there is tunneling between layers which are separated because of strong $B_{||}$ (see Fig. 1(E) inset). 

We acknowledge support through the NSF (Grants DMR-1305691, ECCS-1508925 and MRSEC DMR-1420541), the Gordon and Betty Moore Foundation (Grant GBMF4420), and the Keck Foundation for sample fabrication and characterization, the DOE BES (Grant DE-FG02-00-ER45841) for measurements and the NSF (Grant DMR-1310199) for calculations. This work was performed at the National High Magnetic Field Laboratory, which is supported by NSF Cooperative Agreement DMR-1157490, by the State of Florida, and by the DOE. Work at Argonne was supported by DOE BES under Contract DE-AC02-06CH11357. We thank S. Hannahs, T. Murphy, A. Suslov, J. Park and G. Jones at NHMFL for valuable technical support.


\begin{thebibliography}{99}
\bibitem{footnote1}The calculations for $B_{||}=0$ predict a slight ($<\sim{5}\%$) occupation of the upper subband, which we do not observe experimentally.
\bibitem{footnote1_1}The 7.08 T peak is observed because the $B_{\perp}$-range used in the Fourier analysis ($0.2\lesssim{B_{\perp}}\lesssim0.9$ T) includes spin-resolved SdH oscillations starting at ${B_{\perp}}\gtrsim0.5$ T.
\end{thebibliography}
\end{document}